\documentstyle[12pt,psfig]{article}
\begin{document}
\hyphenation{HERWIG}
\hyphenation{JETRAD}
\hyphenation{BFKL}
\begin{titlepage}
\vspace{-2.5cm}         
\begin{flushright}      
{\normalsize FERMILAB-CONF-96/162-E}        
\end{flushright}        

\begin{center}
{\large\bf Jet Decorrelation and Jet Shapes at the Tevatron
\footnotemark           
\footnotetext{Presented at Les Rencontres de                    
              Physique de la Vallee d'Aoste:                    
              Results and Perspectives in Particle Physics,     
              La Thuile, Italy, 3--9 March 1996.}               
} \\
\vspace{2.5cm}
{\large Terry C. Heuring} \\
\vspace{.5cm}
{\sl Florida State University\\ (for the D\O\ Collaboration) }\\
\vspace{2.5cm}
\vfil
\begin{abstract}

We present results on measurements of jet shapes and jet azimuthal decorrelation
from $\overline{p}p$ collisions at $\sqrt{s}=1.8$ TeV using data collected
during the 1992--1993 run of the Fermilab Tevatron.  Jets are seen to narrow
both with increasing $E_T$ and increasing rapidity.  While HERWIG, a parton
shower Monte Carlo, predicts slightly narrower jets, it describes the trend of
the data well; NLO QCD describes qualitative features of the data but is
sensitive to both renormalization scale and jet definitions.  Jet azimuthal
decorrelation has been measured out to five units of pseudorapidity.  While
next--to--leading order QCD and a leading--log approximation based on BFKL
resummation fail to reproduce the effect, HERWIG describes the data well.

\end{abstract}

\end{center}
\end{titlepage}

\section{Introduction}

As progress has extended perturbative quantum chromodynamics (QCD) beyond
leading order, theoretical comparisons of a more detailed nature are being
performed with experimental data.  The advent of both next--to--leading order
(NLO) and leading--log approximations (LLA) has enabled predictions for both
event topologies involving more than two jets and the internal structure within
a jet. This paper reports on two such studies, one involving a detailed
examination of the transverse energy flow within jets and the other involving
the first study of azimuthal decorrelation of jets with large separations in
pseudorapidity.

In both studies, NLO QCD is the first order in which a theoretical prediction is
possible.  Although NLO predictions have been successful in describing inclusive
single jet and dijet~\cite{jet 1} cross sections, deviations from 
the measured distributions or a large sensitivity to the choice of
renormalization scale may indicate large higher order corrections.   In this
case, an all--orders approximation in the form of a LLA may produce better
results.  For the jet shape analysis, a model where only one additional radiated
parton is allowed may prove inadequate  to reproduce all aspects of the
transverse energy flow.  Additional fragmentation effects may indeed be
important.  In jet production with large rapidity separations, the presence of
multiple scales in the calculation may require a resummation.  Two alternative
methods are explored, one resumming $\ln Q^2$ via DGLAP~\cite{dglap} splitting
functions and the other resumming $\ln(\hat{s}/Q^2)$ using the BFKL~\cite{bfkl}
equation.  A comparison of experimental results to theoretical predictions will
determine the necessity of including such higher order corrections.

\section{Jet Shape}

The data for this study~\cite{jet shape} were collected using the D\O\ detector
which is described in detail elsewhere~\cite{detector}.  The critical components
for this study are the uranium--liquid argon calorimeters.  These calorimeters
provide hermetic coverage over most of the solid angle with a transverse
segmentation of $0.1 \times 0.1$ in $\eta \times \phi$ ($\eta =
-\ln[\tan(\theta/2)]$, where $\theta$ is the polar angle with respect to the
proton beam and $\phi$ is the azimuthal angle).  The measured resolutions are
$15\%/\sqrt{E}$ and $50\%/\sqrt{E}$ (E in GeV) for electromagnetic and single
hadronic showers respectively.

The data were collected using four separate triggers, each requiring a minimum
$E_{T}$ in a specified number of trigger towers ($0.2 \times 0.2$ in $\eta
\times \phi$) at the hardware level.  Events satisfying these requirements were
sent to a processor farm where a fast version of the offline jet algorithm
searched for jet candidates above some $E_{T}$ threshold.  These events were
used to populate four non-overlapping jet $E_{T}$ ranges: 45-70, 70-105,
105-140, and $> 140$ GeV.  Each trigger was used only where it was
fully efficient.

Jets were reconstructed offline using an iterative fixed cone algorithm with a
radius R = 1.0 in $\eta - \phi$ space.  The final jet angles were defined as
follows: $\theta_{\rm jet} = \tan^{-1} \sqrt{(\Sigma_i E_{xi})^2 + (\Sigma_i
E_{yi})^2)}/\Sigma_i E_{zi}$ and $\phi_{\rm jet} = \tan^{-1}(\Sigma_i E_{yi}
/ \Sigma_i E_{xi})$ with $z$ in the beam direction and the sums extending over
calorimeter towers whose centers were contained within the jet cone.  These
definitions differ from the Snowmass algorithm~\cite{Snowmass}: $\eta_{\rm jet}
= \Sigma_i E_{Ti}\eta_i / \Sigma_i E_{Ti}$ and $\phi_{\rm jet} = \Sigma_i E_{Ti}
\phi_i / \Sigma_i E_{Ti}$.  In both cases, the $E_T$ of the jet was defined as
$\Sigma_i E_{Ti}$.  After a preliminary set of jets was found, overlapping jets
were redefined.  Two jets were merged into one jet if more than 50\% of the
$E_{T}$ of the jet with the smaller $E_{T}$ was contained in the overlap region.
The direction of the new jet was defined as the vector sum of the two original
jet momenta, and the energy was recalculated.  If less than 50\% of the $E_{T}$
was contained in the overlap region, the jets were split into two distinct jets.
In this case, the energy of each calorimeter cell in the overlap region was
assigned to the nearest jet and the jet directions were recalculated.  Two
different $\eta$ regions ($|\eta| < 0.2$ and $2.5 < |\eta| < 3.0$) were studied.

For this study we define the quantity $\rho(r) = (1/N_{\rm jets}) 
\Sigma_{\rm jets} E_{T}(r)/E_{T}(r=1)$ ($N_{\rm jets}$ is the number of jets
in the sample) which is the average fraction of the jet transverse energy
contained in a subcone of radius $r$.  For a given value of $r$, a larger value
of $\rho$ represents a narrower jet. Individual subcones were corrected for
underlying event and noise effects only.

\begin{figure}
 \centerline{
      \psfig{figure=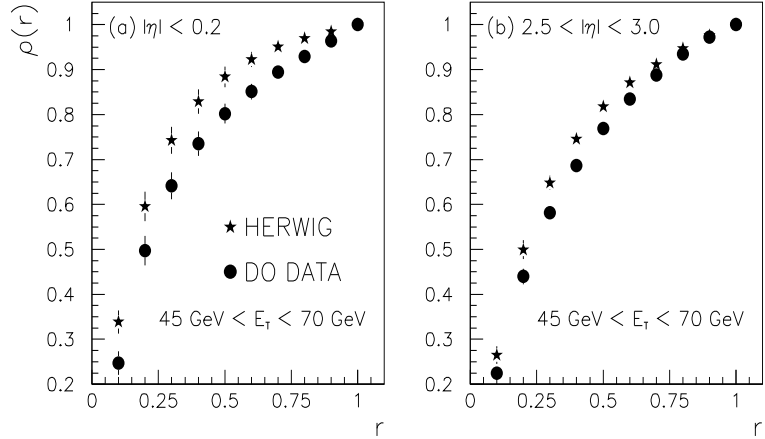,width=4.5in,height=3.0in}}
 \caption{The average integrated $E_T$ fraction versus the subcone radius is
 plotted for the data and HERWIG Monte Carlo, before calorimetric effects are
 removed, for the $E_T$ range 45-70 GeV for (a) $|\eta| \le 0.2$ and (b) $2.5
 \le |\eta| \le 3.0$.}
 \label{js 1}
\end{figure}

In Fig.~\ref{js 1}, the average $E_T$ fraction ($\rho$) versus the subcone
radius is shown for the two different $\eta$ regions for jets between 45 and 70
GeV $E_T$.  Also shown in this figure are the predictions from
HERWIG~\cite{herwig}, a Monte Carlo prediction which incorporates higher order
effects through a parton shower model.  Detector effects were modeled using a 
detailed detector simulation based on GEANT~\cite{geant}.  Although properly
describing the qualitative behavior of the data, in both the central
and forward regions the theoretical predictions are narrower than the measured
jet shapes.  In order to remove detector effects, correction factors were
determined using three different event generators, comparing the jet shape
before and after detector simulation.  In the subsequent distributions, the
detector effects have been removed using these correction factors.

\begin{figure}
 \centerline{
      \psfig{figure=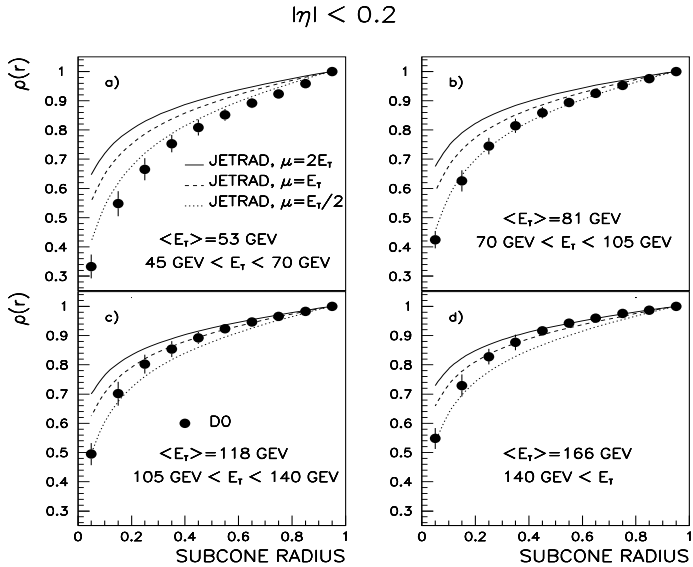,width=4.0in,height=3.5in}}
 \caption{The measured jet shapes, with calorimetric effects  removed, compared
 to NLO predictions (JETRAD) with three renormalization scales for $|\eta| \le
 0.2$ for the jet $E_T$ range (a) 45-70 GeV, (b) 70-105 GeV, (c) 105-140 GeV,
 (d) greater than 140 GeV.}
 \label{js 2}
\end{figure}

\begin{figure}
  \centerline{\psfig{figure=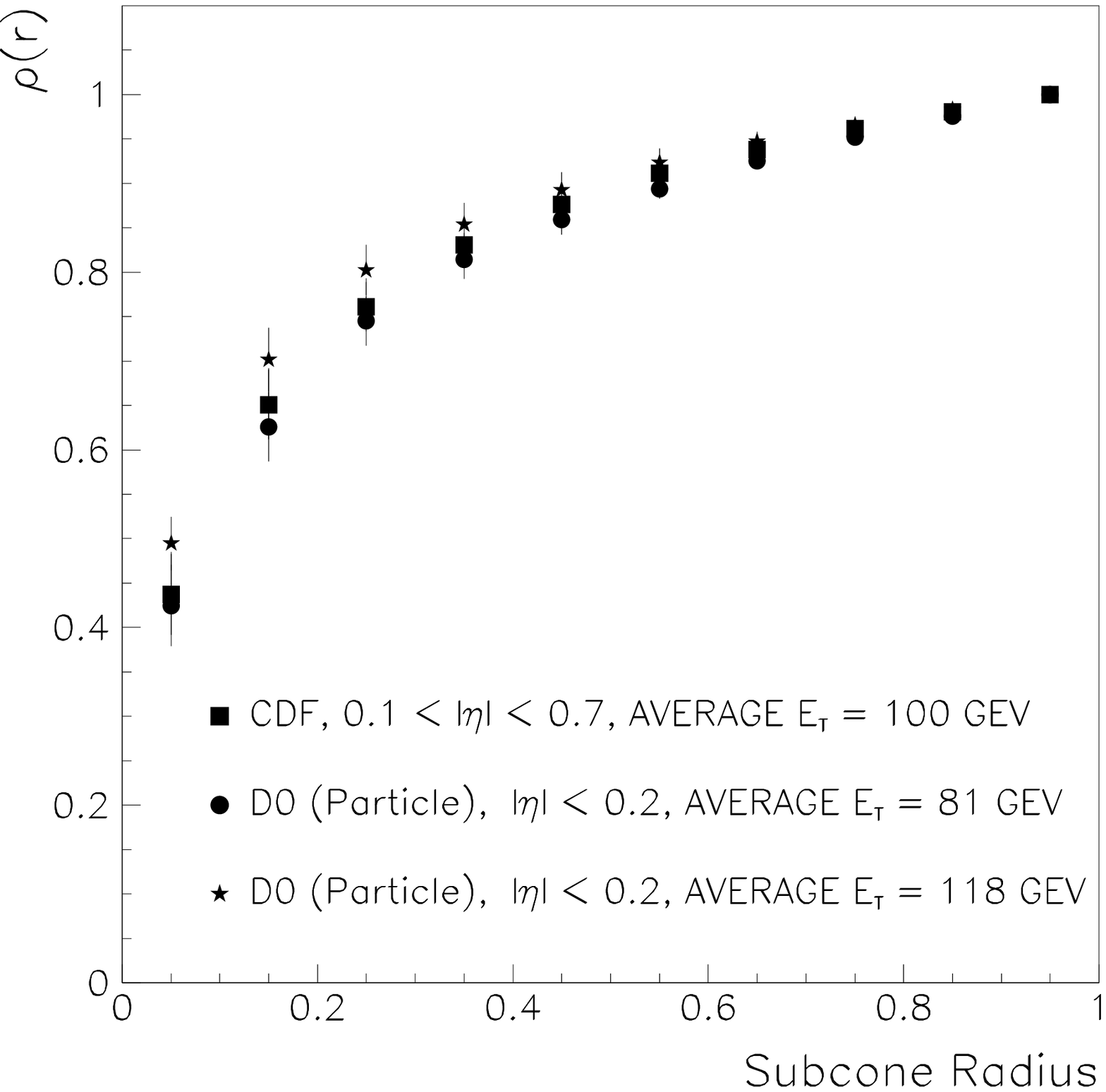,width=3.0in}}
  \caption{A comparison of the D\O\ and CDF jet shapes.  The CDF measurement was
  made using charged tracks only for $0.1 < |\eta| < 0.7$ and average $E_T =
  100$ GeV.  The two D\O\ measurements ($|\eta| \le 0.2$), with detector effects
  removed, were made at $E_T$'s above and below that of CDF.  Good agreement is
  seen.}
  \label{js 4}
\end{figure}

\begin{figure}
  \centerline{\psfig{figure=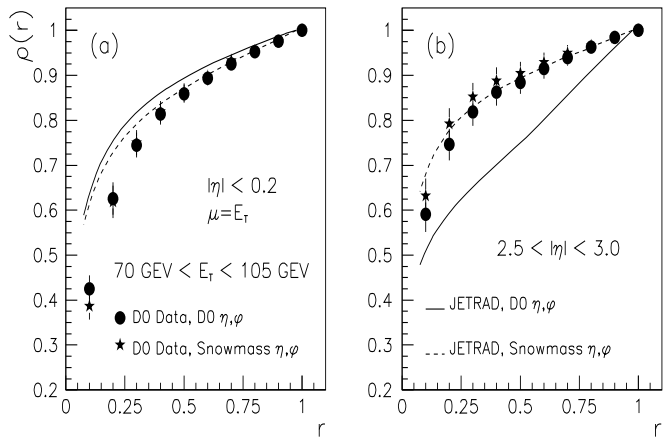,width=4.0in,height=3.0in}}
  \caption{Comparisons of jet shapes from data and NLO predictions using
  different jet direction definitions for jets with $70 < E_T < 105$ and for (a)
  $|\eta| \le 0.2$ and (b) $2.5 \le |\eta| \le 3.0$.}
  \label{js 3}
\end{figure}

In Fig.~\ref{js 2}, the jet shape for the four different $E_{T}$ regions are
shown for central jets ($|\eta| < 0.2$).  We see that as the $E_T$ of the jet
increases, a larger fraction of its total $E_T$ is contained in the inner
subcones, i.e. the jets are narrower.  This result is in good agreement with 
CDF~\cite{cdf} measurements (Fig. \ref{js 4}) where jet shapes were determined
using charged particle tracks only.   In Fig.~\ref{js 3}, the jet shape for a
single $E_T$ range in both the central and the forward $\eta$ region is shown.
We see that jets of the same $E_T$ become narrower as the jets become more
forward.


Also shown in Figs.~\ref{js 2} and \ref{js 3} is a comparison to NLO QCD
predictions made with JETRAD~\cite{jetrad} using
CTEQ2M~\cite{cteq} parton distribution functions (pdf) and three different
renormalization scales($\mu$): $2E_T$, $E_T$, and $E_T/2$.
Studies showed the theoretical predictions were insensitive to choices in pdf.
Since this is a NLO QCD prediction containing at most three partons in the final
state, jet shape arises from combining two final state partons into one jet.
For this study, two partons were considered to constitute one jet if their
distance in $\eta$ -- $\phi$ space from their vector sum was less than one.
Using the D\O\ definitions of jet angles, central jets are found to be
generally wider than predicted by JETRAD.  Forward jets are narrower than
predicted and the observed narrowing of forward jets with increasing $E_T$ is
not reproduced by JETRAD.

As shown in Fig.~\ref{js 3}, using the Snowmass definition of jet angles
generally improves the agreement between data and predictions.  While the
experimental jet shape distribution is only moderately affected by the choice of
jet angle definition, the theoretical prediction in the forward region  is very
sensitive to this choice.  Qualitatively, the NLO calculations using the
Snowmass algorithm describe the measured jet shapes reasonably well.  However,
these predictions, being the first order in which a nontrivial jet shape
emerges, show large sensitivity to the choice of renormalization scale with no
single choice able to describe the data at all $E_T$'s.

\section{Jet Azimuthal Decorrelation}
 
In this study~\cite{decorrelation} we are interested in jets widely separated in
pseudorapidity and at relatively low $E_T$.  By measuring the $\phi$ correlation
of the dijet system we gain information about the radiation pattern in the
event.  The data used for this study were collected in a similar fashion to
the jet shape study.  In this case, a single trigger was used.  It required
$E_T > 7$ GeV at the hardware level and searched for jet candidates with an
$E_T$ threshold of 30 GeV in the processor farm.  Offline, jets were
reconstructed using an iterative fixed cone algorithm with a radius of R = 0.7
in $\eta - \phi$ space.  Both the final jet angles and the merging and splitting
criteria were defined as described above.

From the sample of jets collected with $E_T > 20$ GeV and $|\eta| < 3.0$, the
two jets at the extremes of pseudorapidity were selected.  To remove any
trigger inefficiencies, one of these two jets was required to have $E_T > 50$
GeV.  The relevant quantities used in this analysis are $\Delta\eta = |\eta_1 -
\eta_2|$ and $\Delta\phi = \phi_1 - \phi_2$ (where the subscripts refer to the
two jets described above).  

\begin{figure}
 \vspace{-0.5in}
 \begin{center}
 \begin{minipage}[t]{2.6in}
  \centerline{\psfig{figure=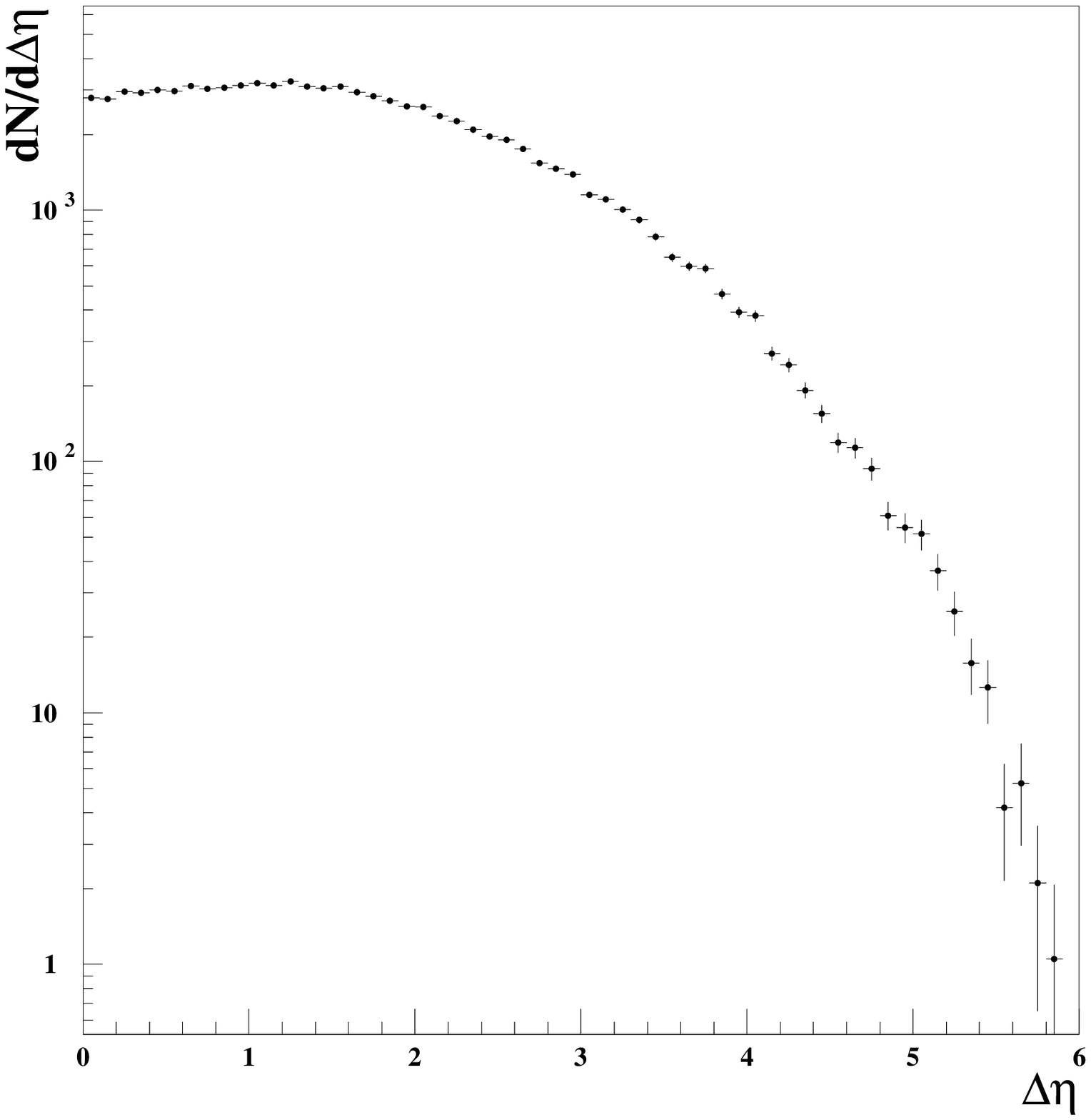,width=2.6in}}
  \caption{The pseudorapidity interval, $\Delta\eta = |\eta_1 - \eta_2|$, of the
  two jets at the extremes of pseudorapidity.  The coverage extends to
  $\Delta\eta \sim 6$.  The errors are statistical only.}
  \label{jd 1}
 \end{minipage} \hspace{.2in}
 \begin{minipage}[t]{2.6in}
  \centerline{\psfig{figure=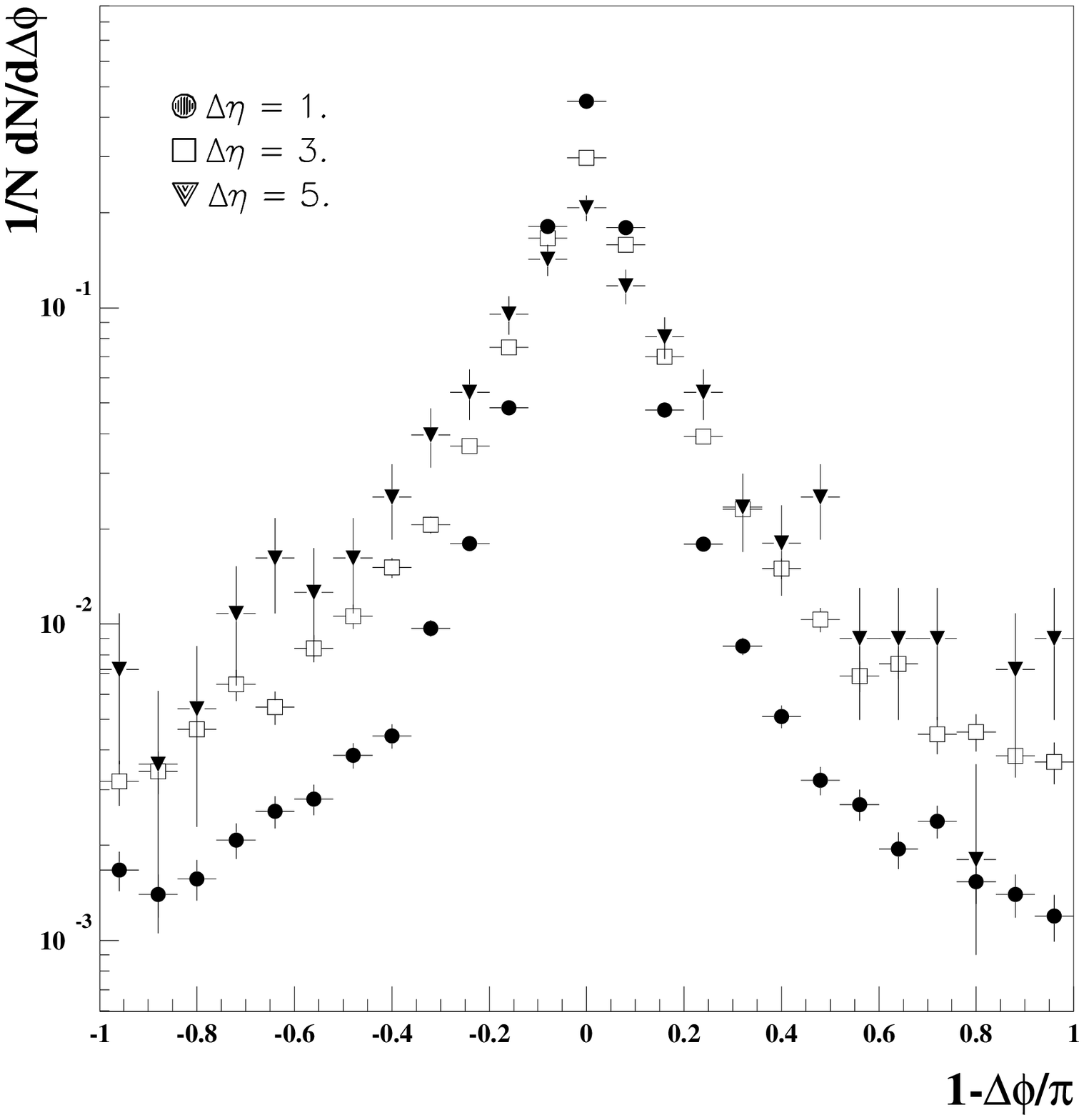,width=2.6in}}
  \caption{The azimuthal angle difference, $\Delta\phi = \phi_1 - \phi_2$,
  distribution of the two jets at the extremes of pseudorapidity plotted as $1
  - \Delta\phi/\pi$ for $\Delta\eta =$ 1, 3, and 5.  The errors are statistical
  only.}
  \label{jd 2}
 \end{minipage}
 \end{center}
\end{figure}

\begin{figure}
 \vspace{-0.5in}
 \centerline{\psfig{figure=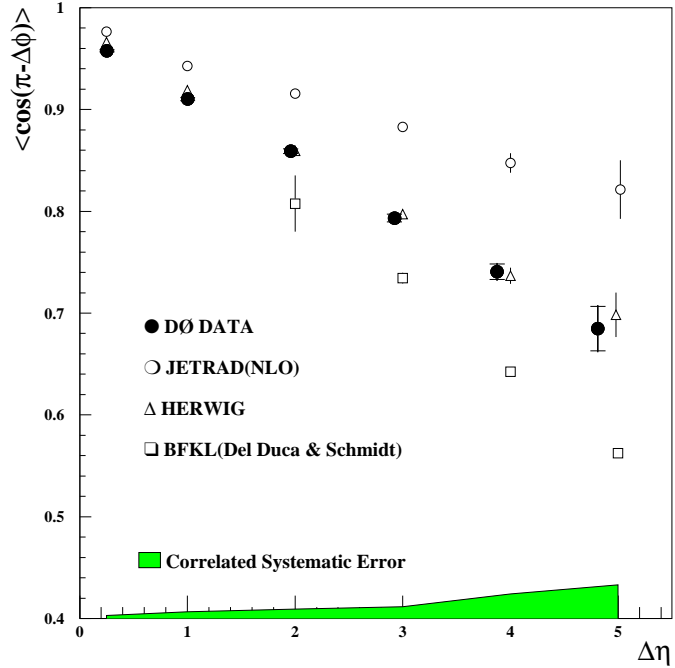,width=4in}}
 \caption{The correlation variable used in this analysis,
 $\langle\cos(\pi-\Delta\phi)\rangle$ vs. $\Delta\eta$, for the data, JETRAD,
 HERWIG, and the BFKL calculations of Del Duca and Schmidt.}
 \label{jd 3}
\end{figure}

Since the trigger was implemented out to $|\eta| < 3.0$, the maximum value
of $\Delta\eta$ is $\sim 6.0$.  This is illustrated in Fig.~\ref{jd 1} where we
plot the pseudorapidity interval $\Delta\eta$ for events passing our cuts.  In
Fig.~\ref{jd 2}, the azimuthal angular separation, $1 - \Delta\phi/\pi$ is
plotted for three selected unit bins of $\Delta\eta$ centered at $\Delta\eta =$
1, 3, and 5.  The correlation of the two jets at the extremes in pseudorapidity
is evident in the shape of this distribution which becomes wider (less
correlated) at $\Delta\eta$ increases.  A correlation variable,
$\langle\cos(\pi-\Delta\phi) \rangle$, has been defined to study the effect
quantitatively.  This quantity is plotted in Fig.~\ref{jd 3} as a function of
$\Delta\eta$.  For the data, the error bars represent the statistical and
uncorrelated systematic errors added in quadrature.  An error band at the bottom
of the plot represents the correlated systematic error due to the energy scale
and detector effects. Also shown in Fig.~\ref{jd 3} are the theoretical
predictions from HERWIG, NLO QCD as implemented in JETRAD, and BFKL
resummation~\cite{DDS}.  The errors for the predictions are statistical only.

For the data, many sources of systematic error were investigated.  For all
values of $\Delta\eta$ the energy scale error dominates, except for $\Delta\eta
= 5$, where the measurement is statistically limited.  Other systematic errors
due to the out--of--cone showering correction, angular biases present in the
jet reconstruction, and our selection cuts are included.  Although no attempt
was made to correct back for detector effects, the size of any such
effects was estimated using HERWIG events that were subjected to a detailed
simulation of our detector based on GEANT.   The size of the effect
was negligible for $\Delta\eta \le 3$ and only $\sim 0.03$ at $\Delta\eta = 5$
and are included in the systematic correlated error band. 

Systematic effects were also studied in the theory.  The renormalization and
factorization scale $\mu$ was varied in the NLO prediction from $\mu =
p_T^{max}/2$ to $2p_T^{max}$.  The largest variation seen in
$\langle\cos(\pi-\Delta\phi)\rangle$ was less than 0.026.  Different parton
distribution functions were used (CTEQ2M, MRSD$-$~\cite{mrs}, and
GRV~\cite{grv}) resulting in variations of less than 0.003.  In addition, the
Monte Carlo study was repeated using the Snowmass jet angle definitions.  The
difference in $\langle\cos(\pi-\Delta\phi)\rangle$ was less than 0.013.  As a
cross check, a subset of the data was analyzed using the Snowmass jet angle
definitions as well.  The resulting variations were less than 0.002.

The data in Fig.~\ref{jd 3} show a nearly linear decorrelation effect with
pseudorapidity.  For small $\Delta\eta$ both HERWIG and JETRAD describe the
data well.  However, at large $\Delta\eta$ JETRAD, which is leading order in
describing any decorrelation effect, shows deviations from the measurement
predicting too little decorrelation.  The BFKL predictions, valid for large
$\alpha_s\Delta\eta$, are shown for $\Delta\eta \ge 2$.  As the pseudorapidity
interval increases, this leading--log approximation predicts too much
decorrelation.  The HERWIG prediction, which includes higher order processes in
the form of a parton shower model including angular ordering, describes the data
well over the entire $\Delta\eta$ region.

\section{Conclusion}

We have presented results on both jet shapes and azimuthal decorrelation in
dijet systems.  These results have been compared to various theoretical
predictions.  We see that jets become narrower as their $E_T$ increases for a
fixed $\eta$ and that they become narrower as their $\eta$ increases for fixed
$E_T$.  The measured jet shapes are insensitive to jet angle definitions while
the NLO QCD predictions are sensitive to jet angle definition.  Reasonable
behavior is seen when Snowmass jet angle definitions are used.  We see that the
dijet systems decorrelate in a linear fashion out to $\Delta\eta = 5$.
Comparisons to various theoretical predictions show that NLO QCD predicts too
little while BFKL resummation predicts too much decorrelation. HERWIG reproduces
the data over the entire $\Delta\eta$ range explored.

\end{document}